\documentclass[pdflatex,sn-mathphys-num]{sn-jnl}% Math and Physical Sciences Numbered Reference Style

\usepackage{graphicx}%
\usepackage{multirow}%
\usepackage{amsmath,amssymb,amsfonts}%
\usepackage{amsthm}%
\usepackage{mathrsfs}%
\usepackage[title]{appendix}%
\usepackage{xcolor}%
\usepackage{textcomp}%
\usepackage{manyfoot}%
\usepackage{booktabs}%
\usepackage{algorithm}%
\usepackage{algorithmicx}%
\usepackage{algpseudocode}%
\usepackage{listings}%
\usepackage{subfig}% used for the two-panel star-graph figure (Fig.~\ref{fig:star})
\begin{document}

\title[Discrete Dirac equation on quantum graphs]{Discrete Dirac equation on quantum graphs: A model of a Dirac particle in a branched lattice}

\author[1]{\fnm{M.} \sur{Akramov}}\email{mashrabresearcher@gmail.com}

\author*[2]{\fnm{C.} \sur{Trunk}}\email{carsten.trunk@tu-ilmenau.de}

\author[3]{\fnm{D.} \sur{Matrasulov}}\email{dmatrasulov@gmail.com}

\affil*[1]{\orgname{National University of Uzbekistan}, \orgaddress{\street{Universitet Str. 4}, \city{Tashkent}, \postcode{100174}, \country{Uzbekistan}}}

\affil[2]{\orgdiv{Institute for Mathematics}, \orgname{TU Ilmenau}, \orgaddress{\street{Weimarer Str. 25}, \city{Ilmenau}, \postcode{98693}, \country{Germany}}}

\affil[3]{\orgname{Turin Polytechnic University in Tashkent}, \orgaddress{\street{Niyazov Str. 17}, \city{Tashkent}, \postcode{100095}, \country{Uzbekistan}}
\\
\ 
\\
\ 
\\
Dedicated to the memory of 
Franciszek Hugon Szafraniec, a great mathematician whose work and personality have been a lasting source of inspiration.}

\abstract{We study the one-dimensional time-independent discrete Dirac equation on graphs with discrete edges. Both continuous and discrete formulations of the graphs are considered, with particular focus on star graphs with three edges as a simple example. Exact solutions for eigenstates and spectra are derived and compared with the continuum case, showing good agreement in tabulated results. This approach offers a systematic framework for analyzing relativistic quantum transport on branched structures relevant to nanostructures and quantum device design.}


\maketitle

\section{Introduction}\label{sec1}

Modeling quantum particle dynamics in complex branched networks is highly relevant for advancing a range of technological applications, particularly in engineering quantum devices with tailored properties. An effective theoretical framework for this is provided by quantum graph models. Their strength lies in simplifying the analysis of transport phenomena in branched low-dimensional systems by reducing the problem to solving one-dimensional Dirac equations defined on discrete graph structures. In many cases, this approach enables precise solutions even for graphs with arbitrary connectivity and branching patterns.

Earlier studies addressed the confinement of a relativistic particle in a one-dimensional potential well, demonstrating the need for alternative boundary conditions to obtain meaningful solutions~\cite{Alonso1, Alonso2}. Dirac and Weyl fermions have also been widely studied in condensed matter systems---from low-energy excitations in Dirac and Weyl materials to quantum criticality in 2D systems and confined dynamics with time-dependent boundaries---highlighting their rich physical properties and experimental signatures (see Refs.~\cite{d_material, Boyack, Hasan, Rakhmanov}).

Various models and methods have been developed to study quantum transport, spectral properties, boundary conditions, and inverse problems on graphs and molecular systems, highlighting connections between geometry, spectrum, and physical observables (see Refs.~\cite{Pauling,Rud,Alex,Exner1,Kurasov}). In Refs.~\cite{Uzy1, Kuchment04, Uzy2, Gaspard, Exner15, Grisha, Barra, Uzy3, Mugnolo, Uzy4, Bolte1, Hul}, quantum graphs have been extensively studied to explore spectral statistics, trace formulas, boundary conditions, resonance phenomena, and experimental realizations, revealing rich connections to quantum chaos and random matrix theory.

Following the work of Kostrykin and Schrader~\cite{Kost} on the Schr\"odinger operator on graphs, Bolte and Harrison~\cite{Bolte} investigated the spectral statistical properties of the Dirac equation on graphs. Spin effects in the spectral statistics of quantum graphs with Dirac operators have been analyzed in Ref.~\cite{Bolte2}, while spectral zeta functions and determinants for Dirac operators on metric graphs with general matching conditions have been developed in Ref.~\cite{Harrison}.

Dirac particles on periodic~\cite{Dirac_tbc2} and transparent~\cite{Dirac_tbc1} quantum graphs have been studied by applying continuity and current conservation conditions at the graph vertices. Dirac operators with singular interactions~\cite{Behrndt} and on noncompact metric graphs~\cite{Borrelli} have been analyzed using boundary triplet methods to study self-adjointness and spectral properties. The Dirac operator on composite one-dimensional structures with half-lines and intervals has been studied via junction conditions linking parameters to spectral data (see Ref.~\cite{Bulla}).

Spectrum of a nonselfadjoint discrete Dirac operator with general boundary conditions has been studied in Ref.~\cite{DDE}. In Ref.~\cite{DDE2}, a projection method was developed to discretize the Klein--Gordon and Dirac equations while preserving the momentum operator's form and the unitary Fourier transform properties. As shown in Ref.~\cite{DDE3}, discrete Dirac operators on a square lattice converge strongly to the continuous Dirac operators as the mesh size tends to zero.

From an operator-theoretic point of view, discrete quantum models are closely related to the spectral analysis of unbounded operators, Jacobi-type operators, and operator realizations of fundamental quantum-mechanical observables. Important contributions in this direction were made by F.H.\ Szafraniec and his collaborators, including operator-theoretic studies of the quantum harmonic oscillator, creation and annihilation operators, Hamiltonians, \(q\)-oscillators, and spectral properties of unbounded operators \cite{S0,S1,S2,S3,S4,S5,S6}. These works illustrate how operator-theoretic methods provide a rigorous framework for the analysis of discrete and continuous quantum systems.

In this paper, we investigate the one-dimensional discrete Dirac equation on graphs with discrete edges, following the approach presented in Ref.~\cite{DSE}, where each edge corresponds to a finite discrete segment supporting Dirac dynamics, effectively modeling quantum transport in branched lattice systems. These structures arise in various areas of contemporary condensed matter physics, including graphene-based nanostructures, topological phases of matter, one-dimensional quantum wires, and conducting polymers. By solving the discrete Dirac equation on a finite one-dimensional chain, we build solutions on the discrete quantum graph that obey boundary conditions at the vertices, with the eigenvalues eventually found from a corresponding secular equation.

This paper is structured as follows. In Section~\ref{section1}, we briefly present the Dirac equation on the finite interval. Section~\ref{dse_on_aline} addresses the discrete Dirac equation on a finite chain. Section~\ref{sec_qg_theory} provides a brief overview of quantum graph theory using the continuum Dirac equation. In Section~\ref{sec_branched}, we present solutions of the discrete Dirac equation on quantum graphs, considering both star-like and general branching structures. Finally, Section~\ref{sec_conclusion} offers concluding remarks.

\section{Dirac equation on a finite interval}\label{section1}
In this section we briefly recall the one-dimensional time-independent Dirac equation on a finite interval $[0,L]$, where $\hbar=c=1$. The equation for a free particle with $V(x)=0$ is given by~\cite{Bolte}
\begin{eqnarray}\label{dirac_line}
\begin{matrix}
m\phi(x)-\partial_x\chi(x) = \varepsilon \phi(x), \\
\partial_x \phi(x)-m\chi(x) = \varepsilon \chi(x),
\end{matrix} \quad
x\in [0,L].
\end{eqnarray}
This formulation describes a relativistic particle in one dimension and leads to a coupled system of first-order differential equations for the components $\phi(x)$ and $\chi(x)$.
Together with \eqref{dirac_line}, we consider the boundary conditions given by
\begin{eqnarray}\label{bc_cont_line}
    \phi(x)|_{x=0}=0 \text{  and  } \phi(x)|_{x=L} = 0.
\end{eqnarray}

Note that the Dirac operator associated with \eqref{dirac_line} and \eqref{bc_cont_line} defined on
$H^1([0,L],\mathbb{C}^2)$, see, e.g. \cite[Chapter 15]{Weidmann1}.
The solution of~\eqref{dirac_line} can be written explicitly as
\begin{eqnarray}\label{sol_cont}
    \begin{pmatrix}
        \phi(x) \\ \chi(x)
    \end{pmatrix} = A\begin{pmatrix}
        \sin(k x) \\
        \gamma \cos(k x)
    \end{pmatrix},
\end{eqnarray}
where $A\in\mathbb{C}$ is an arbitrary constant and
\begin{eqnarray}
    \gamma = \frac{\varepsilon-m}{k}, \quad
    \text{and} \quad
    \varepsilon = \sqrt{k^2+m^2}.
\end{eqnarray}
The quantization of the momentum arises from the boundary conditions, leading to the eigenvalues
\begin{eqnarray}
    k = \frac{\pi {\ell}}{L},
\end{eqnarray}
where $\ell\in \mathbb{Z}$.
These eigenvalues describe the relativistic energy levels of the particle confined on the finite interval $[0,L]$, showing the characteristic dependence on mass and momentum in the Dirac formalism.

\section{Discrete Dirac equation in a finite chain}\label{dse_on_aline}

In this section, we consider the one-dimensional time-independent discrete Dirac equation for a free particle by discretizing the finite interval $[0, L]$ as
\begin{eqnarray}
x_n^{(a)}=na, \;\;
\text{where} \;\; n=0,\dots, N \;\;
\text{and} \;\; a \;\;
\text{is the step size.}
\end{eqnarray}
Then the discrete Dirac equation is written as
\begin{gather}\label{discrete_line}
\begin{matrix}
m\phi^{(a)}(x_n^{(a)})-\frac{1}{a} ( \chi^{(a)}(x_{n+1}^{(a)}) - \chi^{(a)}(x_n^{(a)})) = \varepsilon\phi^{(a)}(x_n^{(a)}), \\[1ex]
\frac{1}{a} ( \phi^{(a)}(x_n^{(a)}) - \phi^{(a)}(x_{n-1}^{(a)}))-m\chi^{(a)}(x_n^{(a)}) = \varepsilon\chi^{(a)}(x_n^{(a)})
\end{matrix}, \quad
n=1,...,N-1.
\end{gather}

The discrete Dirichlet boundary conditions are
\begin{eqnarray}\label{dirichlet}
    \phi^{(a)}(x_n^{(a)})\big|_{x_n^{(a)}=0}
    , \quad
    \phi^{(a)}(x^{(a)}_{n})\big|_{x^{(a)}_{n}=x^{(a)}_{N-1}}=0.
\end{eqnarray}

With these boundary conditions, the discrete Dirac equation~\eqref{discrete_line} admits an exact analytical solution of the form
\begin{eqnarray}\label{sol_line2}
    \begin{pmatrix}
    \phi^{(a)}(x^{(a)}_n) \\ \chi^{(a)}(x^{(a)}_n)
    \end{pmatrix} =
    A\begin{pmatrix}
        \sin[g(a) x^{(a)}_n ] \\
        \gamma \cos[ g(a)x^{(a)}_n-r(a) ]
    \end{pmatrix}, \quad
    n=0,\dots,N,
\end{eqnarray}
where  $A \in \mathbb{C}$ is an arbitrary complex constant and
$$
\gamma = \frac{\varepsilon-m}{k}, \quad \mbox{and} \quad \varepsilon=\sqrt{k^2+m^2}.
$$
The functions $g(a)$ and $r(a)$ are defined by
\begin{equation*}
    g(a) = \frac{2}{a} \arcsin\left( \frac{k a}{2} \right), \quad
    r(a)=\arcsin\left( \frac{ka}{2}\right ).
\end{equation*}

The energy eigenvalues associated with the discrete solution are given by
\begin{equation}\label{eigenvalue_line}
   \varepsilon_{\ell}=\sqrt{k_{\ell}^2+m^2}, \quad
   k_{\ell}=\frac{2}{a} \sin\left[\frac{\pi {\ell}}{2(N-1)}\right],
\end{equation}
where $\ell \in \mathbb{Z}$.
One may also verify that
\begin{eqnarray}
    \lim_{a\to 0} k_{\ell} = \frac{\pi {\ell}}{L}.
\end{eqnarray}
thus confirming convergence of the discrete model to the continuous one in the limit of vanishing step size.

In Table~\ref{tab:eigenvalues_table}, we present the first five energy eigenvalues $\varepsilon_1,\dots,\varepsilon_5$ corresponding to different values of step size $a$ for $m=1$. The results demonstrate clear convergence of the discrete spectrum to the continuous case as $a \to 0$.

\begin{table}[t!]
\caption{The first five energy eigenvalues for $\varepsilon$ with the continuous case.}\label{tab:eigenvalues_table}
\centering
\begin{tabular}{l*{5}{|c}r}
\hline
  & Continuous & $a=0.1$    & $a=0.01$  & $a=0.001$  & $a=0.0001$ \\
  &  case      & $N=10$      &  $N=100$  &  $N=1000$  & $N=10000$ \\
\hline
1            & 3.2969083  & 3.6140664  & 3.3270336  & 3.2999038  & 3.2972077  \\
2            & 6.3622651  & 6.9131116  & 6.4238984  & 6.3684663  & 6.3628856  \\
3            & 9.4776811  & 10.0498756  & 9.5687799  & 9.4870279  & 9.4786181 \\
4            & 12.6060966 & 12.8945866 & 12.7241402 & 12.6185532 & 12.6073485 \\
5            & 15.7397621 & 15.3534894 & 15.8815058 & 15.7552924 & 15.7413282 \\
\hline
\end{tabular}
\end{table}

To demonstrate the consistency of the solution with its continuous counterpart, we consider the limit of $g(a)$ and $r(a)$ at $a \to 0$ as
\begin{equation}\label{lim}
    \lim_{a\to 0}g(a)=k, \quad \lim_{a\to 0} r(a) = 0,
\end{equation}
At a fixed point $x_n^{(a)}:=X$, we obtain
\begin{gather} \label{limit}
    \lim_{a\to 0}
    \begin{pmatrix}
    \phi^{(a)}(X) \\ \chi^{(a)}(X)
    \end{pmatrix} =
A\begin{pmatrix}
     \sin(kX) \\
     \gamma \cos(kX)
\end{pmatrix},
\end{gather}
which coincides with the solution of the continuous Dirac equation \eqref{sol_cont} on $[0, L]$.

\section{Basic theory of quantum graphs}\label{sec_qg_theory}
\subsection{Arbitrary quantum graphs}\label{cont_graph}
In this subsection, we provide a concise overview of quantum graphs and their spectral properties in the context of the Dirac equation, following the approach presented in Ref.~\cite{Uzy1}. A quantum graph is defined by a set of $V$ vertices, with $V \in \mathbb{N}$, interconnected by $E$ edges, where $E$ is also a natural number. An edge connecting vertices $i$ and $j$ is referred to as the $(i,j)$ edge. Throughout this work, we consider only simple graphs---each pair of vertices is connected by at most one edge, and self-connections (loops), i.e., edges of the form $(i,i)$, are excluded.

The topology of the graph is defined by its $V\times V$ adjacency matrix as
\begin{eqnarray}\label{C}
    C_{i,j}=C_{j,i}=\left\{
\begin{matrix}
    1, & \text{if} \; i \; \text{and} \; j \; \text{are} \; \text{connected,} \\
    0, & \text{if} \; i=j, \\
    0, & \text{otherwise,}
\end{matrix}
    \right.
\end{eqnarray}
where $i,j=1,2,3,..,V$. The number of edges can be found in terms of the adjacency matrix by
\begin{eqnarray}\label{E}
    E=\frac{1}{2}\sum_{i,j=1}^V C_{i,j}.
\end{eqnarray}

Each edge $(i,j)$ of the graph, for which the adjacency matrix $C_{i,j}=1$ and $i<j$ holds, is associated with a coordinate $x_{i,j}$ defined over the interval $[0, L_{i,j}]$, where $L_{i,j}=L_{j,i}$ denotes the length of the corresponding edge. The spinor $\psi=(\phi,\chi)^T$ is represented as a vector function with $E$ components, where each component $\psi_{i,j}(x_{i,j})=(\phi_{i,j}(x_{i,j}),\chi_{i,j}(x_{i,j}))^T$ is defined on the respective edge $(i,j)$ with $C_{i,j}=1$. On each edge, the Dirac equation takes the standard form (assuming units where $\hbar = c = 1$):
\begin{equation}\label{dirac_graph}
\begin{matrix}
m\phi_{i,j}(x_{i,j})-\partial_x\chi_{i,j}(x_{i,j}) = \varepsilon \phi_{i,j}(x_{i,j}), \\
\partial_x \phi_{i,j}(x_{i,j})-m\chi_{i,j}(x_{i,j}) = \varepsilon \chi_{i,j}(x_{i,j}),
\end{matrix} \quad
x_{i,j}\in [0,L_{i,j}].
\end{equation}

In particular, for each vertex there exists a complex number $\varphi_i$ with $i\in \{1, \ldots, V\}$, such that
\begin{equation}\label{bc00}
    \phi_{i,j}(x_{i,j})\big|_{x_{i,j}=0}=\varphi_i, \quad
    \phi_{i,j}(x_{i,j})\big|_{x_{i,j}=L_{i,j}}=\varphi_j,
\end{equation}
for $i,j$ such that $C_{i,j}=1$, $i<j$,
and some current conservation rule~\cite{Uzy1},
\begin{align}\label{bc5}
\begin{split}
    \sum_{j>i} C_{i,j}
    \chi_{i,j}(x_{i,j})\big|_{x_{i,j}=0}
    -\sum_{j<i} C_{i,j} \chi_{j,i}(x_{j,i})\big|_{x_{j,i}=L_{j,i}}
     = \lambda_i \varphi_i,
\end{split}
\end{align}
for some real parameters $\lambda_i$ for $i$ in $\{1,...,V\}$.
Note that the case $\lambda_i=0$ for all  $i \text{ in } \{1,...,V\}$ is called
Kirchhoff rule.

The solution of the Dirac equation on the graph in~\eqref{dirac_graph}, which also satisfies \eqref{bc00}, can be written as
\begin{gather}\label{sol_continuous}
    \begin{pmatrix}
        \phi_{i,j}(x_{i,j}) \\
        \chi_{i,j}(x_{i,j})
    \end{pmatrix}
    = \frac{C_{i,j}}{\sin(kL_{i,j})}
    \left[
    \varphi_j \begin{pmatrix}
        \sin(kx_{i,j}) \\ \gamma \cos(kx_{i,j})
    \end{pmatrix} -
    \varphi_i \begin{pmatrix}
        \sin[k(x_{i,j}-L_{i,j})] \\ \gamma \cos[k(x_{i,j}-L_{i,j})]
    \end{pmatrix}
    \right],
\end{gather}
for all $i<j$, where
$$
\gamma = \frac{\varepsilon-m}{k}\quad \mbox{and}\quad \varepsilon=\sqrt{k^2+m^2}.
$$
Current conservation rule \eqref{bc5} leads to
\begin{align}\label{bc6}
\begin{split}
    \sum_{j=1}^V C_{i,j} \left[
    \varphi_{j} \frac{1}{\sin(kL_{j,i})}-\varphi_{i}\cot(kL_{i,j})
    \right]
    = \frac{ \lambda_i \varphi_i }{\gamma},
\end{split}
\end{align}
for all $i$ in $\{1,...,V\}$.

This forms a system of linear homogeneous equations for the unknown $\varphi_i$, which admits a non-trivial solution only if
\begin{eqnarray}
    \det(M(k))=0,
\end{eqnarray}
where
\begin{multline}
    M(k)=\\
    \begin{pmatrix}
        \frac{\lambda_1}{\gamma}+\sum_{j=1}^V C_{1,j}\cot(kL_{1,j}) & -\frac{C_{1,2}}{\sin(kL_{1,2})} & \cdots & -\frac{C_{1,V}}{\sin(kL_{1,V})} \\
        -\frac{C_{2,1}}{\sin(kL_{2,1})} & \ddots & \ddots & \vdots \\
        \vdots & \ddots & \ddots & -\frac{C_{V-1,V}}{\sin(kL_{V-1,V})} \\[1.5ex]
        -\frac{C_{V,1}}{\sin(kL_{V,1})} & \cdots & -\frac{C_{V,V-1}}{\sin(kL_{V,V-1})} & \frac{\lambda_V}{\gamma}+\sum_{j=1}^V C_{V,j}\cot(kL_{V,j})
    \end{pmatrix}.
\end{multline}

\subsection{Quantum star graph}\label{dqs_graph}

In this subsection, we examine the simplest type of quantum graph consisting of three edges, commonly known as a star graph, see Figure~\ref{fig:star}a for a star graph with four vertices.
For a graph with this topology, the corresponding adjacency matrix is given by
\begin{eqnarray}
    C=\begin{pmatrix}
        0 & 1 & 1 & 1 \\
        1 & 0 & 0 & 0 \\
        1 & 0 & 0 & 0 \\
        1 & 0 & 0 & 0
    \end{pmatrix}.
\end{eqnarray}
The coordinate on each edge is denoted by $x_{1,j}$ and is defined over the interval $[0, L_{1,j}]$ for $j=2,3,4$. The vertex boundary conditions given in~\eqref{bc00} and~\eqref{bc5} apply for each $j=2,3,4$
\begin{align}\label{bc4}
\begin{split}
    & \phi_{1,j}(x_{1,j})\big|_{x_{1,j}=0} = \varphi_1, \;\;
    \sum_{j=2}^4 \chi_{1,j}(x_{1,j}) \big|_{x_{1,j}=0}=0, \\
    & \phi_{1,j}(x_{1,j})\big|_{x_{1,j}=L_{1,j}} = \varphi_j, \;\;
    \chi_{1,j}(x_{1,j})
    \big|_{x_{1,j}=L_{1,j}}=0,
\end{split}
\end{align}
where $\lambda_1=\lambda_2=\lambda_3=\lambda_4=0$. By substituting the solution of the Dirac equation from~\eqref{sol_continuous} for the star graph into the vertex boundary conditions given in~\eqref{bc4}, we obtain the following secular equations
\begin{eqnarray}
    \sum_{j=2}^4 \tan(k L_{1,j}) = 0.
\end{eqnarray}
By computing the discrete values of $k$, we list the first five positive nonzero energy eigenvalues $\varepsilon$ in the first column of Table~\ref{tab:eigenvalues_star}, using the relation $\varepsilon = \sqrt{k^2 + m^2}$.

\section{Discrete Dirac equation on branched lattices}\label{sec_branched}
\subsection{General branching structure}
Following the framework for quantum graphs introduced in~\cite{Uzy1}, this subsection outlines a method for solving the discrete Dirac equation on discrete quantum graphs.

A discrete quantum graph consists of a set of $V$ vertices connected by $E$ discrete edges. Each edge linking vertices $i$ and $j$ (with $i < j$) is denoted by $(i,j)$.

For each $(i,j)$ with $C_{i,j} = 1$ means that vertex $i$ and vertex $j$ are connected.
The topology of the graph is again given by the adjacency matrix \eqref{C} and, hence, $E$ is given by \eqref{E}.
We assign a discrete set of coordinates along the edge: $x_0^{(a_{i,j})}, \ldots, x_{N_{i,j}}^{(a_{i,j})}$, where the spacing between points is given by the step size $a_{i,j}=a_{j,i}$. The left endpoint is defined as $x_0^{(a_{i,j})} = 0$, while the right endpoint is $x_{N_{i,j}}^{(a_{i,j})} = L_{i,j} = N_{i,j} a_{i,j}$. In other words, each edge is treated as a discrete interval that begins at zero and ends at length $L_{i,j}=L_{j,i}$.

The spinor $\psi=(\phi,\chi)^T$ is represented as a vector with $E$ components, where each component corresponds to an edge $(i,j)$ and is given by a vector in $\mathbb{C}^{2N_{i,j}+2}$, defined at the discrete points along that edge,
\begin{equation*}
    \left(
    \begin{matrix}
        \phi_{i,j}^{(a_{i,j})}(x_{n_{i,j}}^{(a_{i,j})}) \\
        \chi_{i,j}^{(a_{i,j})}(x_{n_{i,j}}^{(a_{i,j})})
    \end{matrix}
    \right)_{n_{i,j}=0}^{N_{i,j}},
\end{equation*}
with $C_{i,j}=1$, where $i<j$.

The discrete Dirac equation on each edge is given by (see also \eqref{discrete_line})
\begin{align}\label{eq_graph}
\begin{matrix}
m \phi_{i,j}^{(a_{i,j})}(x_{n_{i,j}}^{(a_{i,j})}) -
\frac{1}{a}\left[\chi_{i,j}^{(a_{i,j})}(x_{n_{i,j}+1}^{(a_{i,j})})-\chi_{i,j}^{(a_{i,j})}(x_{n_{i,j}}^{(a_{i,j})})\right] =
\varepsilon \phi_{i,j}^{(a_{i,j})}(x_{n_{i,j}}^{(a_{i,j})}), \\[1ex]
\frac{1}{a}\left[\phi_{i,j}^{(a_{i,j})}(x_{n_{i,j}}^{(a_{i,j})})-\phi_{i,j}^{(a_{i,j})}(x_{n_{i,j}-1}^{(a_{i,j})})\right]
-m\chi_{i,j}^{(a_{i,j})}(x_{n_{i,j}}^{(a_{i,j})})=
\varepsilon \chi_{i,j}^{(a_{i,j})}(x_{n_{i,j}}^{(a_{i,j})}),
\end{matrix}
\end{align}
where $n_{i,j} = 1,...,N_{i,j}-1.$, $C_{i,j}=1$, $i<j$.

\begin{table}
\caption{The first five nonzero eigenvalues, $\varepsilon$ compared with the continuous case on the star graph.}\label{tab:eigenvalues_star}
\centering
\begin{tabular}{l*{5}{|c}r}
\hline
             & Continuous& $a=0.1$   & $a=0.01$  & $a=0.001$ & $a=0.0001$ \\
\hline
1            & 1.2910053 & 1.3053467 & 1.2924050 & 1.2911449 & 1.2910193 \\
2            & 1.4940273 & 1.5269154 & 1.4972018 & 1.4943436 & 1.4940589 \\
3            & 2.1415725 & 2.2187467 & 2.1493735 & 2.1423530 & 2.1416506 \\
4            & 2.6828561 & 2.7637877 & 2.6911454 & 2.6836866 & 2.6829392 \\
5            & 3.3320407 & 3.0168535 & 3.3465675 & 3.3334968 & 3.3321863 \\
\hline
\end{tabular}
\end{table}

As in Section~\ref{cont_graph}, we impose both a continuity condition and a current conservation rule. The continuity condition is given by
\begin{equation}
    \phi_{i,j}^{(a_{i,j})}(x_{n_{i,j}}^{(a_{i,j})})\Big|_{x_{n_{i,j}}^{(a_{i,j})}=0}=\varphi_i, \quad
    \phi_{i,j}^{(a_{i,j})}(x_{n_{i,j}}^{(a_{i,j})})\Big|_{x_{n_{i,j}}^{(a_{i,j})}=x_{N_{i,j}-1}^{(a_{i,j})}}=\varphi_j,\label{vbc_1}
\end{equation}
for some complex numbers $\varphi_{i}$ for all $i<j$ and
$C_{i,j} = 1$.
The discrete version of the current conservation
\begin{align}
    \sum_{j>i} C_{i,j}
    \chi_{i,j}^{(a_{i,j})}(x_{n_{i,j}}^{(a_{i,j})})\Big|_{x_{n_{i,j}}^{(a_{i,j})}=a_{i,j}} -
    \sum_{j<i} C_{i,j}
    \chi_{j,i}^{(a_{i,j})}(x_{n_{i,j}}^{(a_{i,j})})\Big|_{x_{n_{i,j}}^{(a_{i,j})}=x_{N_{i,j}}^{(a_{i,j})}}
     = \lambda_i \varphi_i, \label{vbc_2}
\end{align}
for some real parameters $\lambda_i$ for $i$ in $\{1,...,V\}$.

Note that there is a difference in the boundary conditions \eqref{vbc_1} and \eqref{vbc_2} compared with those in \eqref{bc00} and \eqref{bc5}. Boundary conditions
\eqref{vbc_1} and \eqref{vbc_2} are chosen in such a way that the discrete Dirac operator is self-adjoint. Here we do not consider the discrete Dirac operator.

Motivated by the structure of the solution in~\eqref{sol_continuous}, we construct an alternative linear combination of the fundamental solutions to satisfy the previously stated continuity and current conservation conditions, in a manner analogous to~\eqref{sol_continuous}. The following function satisfies the discrete Dirac equation~\eqref{eq_graph} on each edge and the continuity condition~\eqref{vbc_1},
\begin{multline}\label{sol_discrete_graph}
    \begin{pmatrix}
        \phi_{i,j}^{(a_{i,j})}(x_{i,j}^{(a_{i,j})}) \\
        \chi_{i,j}^{(a_{i,j})}(x_{i,j}^{(a_{i,j})})
    \end{pmatrix} =
    \frac{C_{i,j}}{\sin[g(a_{i,j})x^{(a_{i,j})}_{N_{i,j}-1}]}  \bigg[
    \varphi_j \begin{pmatrix}
        \sin[g(a_{i,j})x^{(a_{i,j})}_{n_{i,j}}] \\
        \gamma
        \cos[g(a_{i,j})x^{(a_{i,j})}_{n_{i,j}}-r(a_{i,j})]
    \end{pmatrix} \\-
    \varphi_i \begin{pmatrix}
        \sin[g(a_{i,j})(x^{(a_{i,j})}_{n_{i,j}}-x^{(a_{i,j})}_{N_{i,j}-1})] \\ \gamma \cos[g(a_{i,j})(x^{(a_{i,j})}_{n_{i,j}}-x^{(a_{i,j})}_{N_{i,j}-1}) - r(a_{i,j})]
    \end{pmatrix}
    \bigg],
\end{multline}
for $i<j$, where
\begin{equation*}
g(a_{i,j})=\frac{2}{a_{i,j}} \arcsin\left( \frac{k a_{i,j}}{2} \right), \quad
r(a_{i,j})=\arcsin\left( \frac{k a_{i,j}}{2} \right ).
\end{equation*}

Using the limits in~\eqref{lim}, the continuum limit of the solution can be derived by taking the limit $a_{i,j} \to 0$ while keeping the point $x_{i,j}^{(a_{i,j})} := X$ fixed. In a similar manner, as demonstrated in Section~\ref{dse_on_aline}, one obtains:
\begin{align}\label{limit_graph}
    \lim_{a_{i,j}\to 0}
    \begin{pmatrix}
        \phi_{i,j}^{(a_{i,j})}(X) \\
        \chi_{i,j}^{(a_{i,j})}(X)
    \end{pmatrix} =
    \frac{C_{i,j}}{\sin(kL_{i,j})}
    \left[
    \varphi_j \begin{pmatrix}
        \sin(kX) \\ \gamma \cos(kX)
    \end{pmatrix} -
    \varphi_i \begin{pmatrix}
        \sin[k(X-L_{i,j})] \\ \gamma \cos[k(X-L_{i,j})]
    \end{pmatrix}
    \right].
\end{align}
The right-hand side of~\eqref{limit_graph} corresponds to the form of the solution of the continuous Dirac equation on quantum graphs, analogous to the expression given in Equation~\eqref{sol_continuous}.

For the solution in \eqref{sol_discrete_graph} to satisfy the vertex boundary conditions specified in~\eqref{vbc_2} give rise to the following system of homogeneous linear equations:
\begin{multline}
   \sum_{j>i} C_{i,j}
   \frac{\varphi_j \cos[g(a_{i,j})a_{i,j}-r(a_{i,j}) ] - \varphi_i\cos[g(a_{i,j})x^{(a_{i,j})}_{N_{i,j}-2}+r(a_{i,j}) ]}{\sin[g(a_{i,j}) x^{(a_{i,j})}_{N_{i,j}-1}]}
   \\-
   \sum_{j<i} C_{i,j}
   \frac{\varphi_i \cos[g(a_{i,j})x^{(a_{i,j})}_{N_{i,j}}-r(a_{i,j}) ] - \varphi_j\cos[g(a_{i,j})a_{i,j}-r(a_{i,j}) ]}{\sin[g(a_{i,j}) x^{(a_{i,j})}_{N_{i,j}-1}]}
    = \frac{\lambda_i \varphi_i}{\gamma}.\label{secular}
\end{multline}
This is a system of homogeneous equations for $\varphi_i$, which has a non-trivial solution when
\begin{eqnarray}\label{seceq}
    \det[m_{i,j}]\big|_{i,j=1}^V=0,
\end{eqnarray}
where the elements $m_{i,j}$ are defined as
\begin{multline*}
    m_{i,j}=\\
    \Bigg\{ \begin{matrix}
        \frac{\lambda_i}{\gamma} + \sum_{j>i} C_{i,j} \frac{\cos[g(a_{i,j})x^{(a_{i,j})}_{N_{i,j}-2}+r(a_{i,j})]  }{
        \sin[g(a_{i,j})x^{(a_{i,j})}_{N_{i,j}-1} ]} +
        \sum_{j<i} C_{i,j} \frac{\cos[g(a_{i,j})x^{(a_{i,j})}_{N_{i,j}}-r(a_{i,j})]  }{
        \sin[g(a_{i,j})x^{(a_{i,j})}_{N_{i,j}-1} ]}, &  \;\; i=j, \\[1.5ex]
        -C_{i,j} \frac{\cos[g(a_{i,j})a_{i,j}-r(a_{i,j})]  }{
        \sin[g(a_{i,j})x^{(a_{i,j})}_{N_{i,j}-1} ]}, &  \;\; i\neq j.
    \end{matrix}
\end{multline*}

\begin{figure}[t!]
    \centering
    \subfloat[\centering Star graph.]{{\includegraphics[width=85mm]{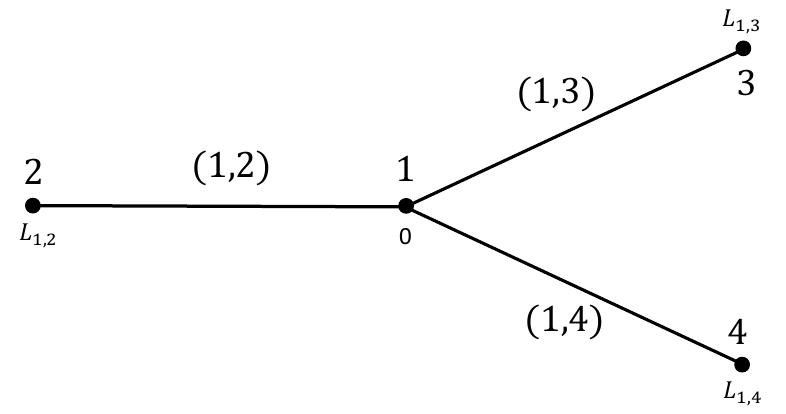} }}%
    \quad
    \subfloat[\centering Discrete star graph.]{{\includegraphics[width=85mm]{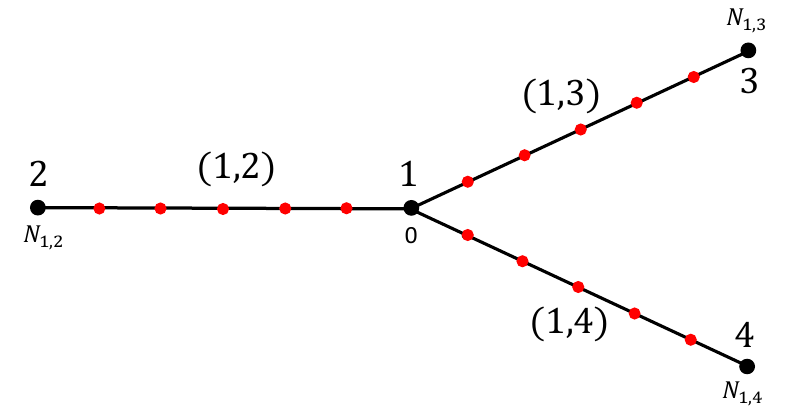} }}
    \caption{Quantum star graphs with four vertices.}
    \label{fig:star}
\end{figure}

\subsection{Star branched lattice}

In this subsection, we analyze a quantum star graph composed of three discrete edges (see Figure~\ref{fig:star}b). For this configuration, the length of each edge $(1,j)$ with $j = 2,3,4$ is given by $N_{1,j} a_{1,j} = L_{1,j}$. The adjacency matrix for this discrete graph is identical to that used in the continuous setting described in the form
\begin{eqnarray}
    C=\begin{pmatrix}
        0 & 1 & 1 & 1 \\
        1 & 0 & 0 & 0 \\
        1 & 0 & 0 & 0 \\
        1 & 0 & 0 & 0
    \end{pmatrix}.
\end{eqnarray}
We set the parameters $\lambda_1=\lambda_2=\lambda_3=\lambda_4=0$ in the boundary condition \eqref{vbc_2}. Under this choice, the continuity conditions in~\eqref{vbc_1} and the current conservation law in~\eqref{vbc_2} apply for each $j = 2,3,4$ as
\begin{align}\label{bc7}
\begin{split}
    & \phi_{1,j}^{(a_{1,j})}(x^{(a_{1,j})}_{n_{1,j}})\Big|_{x^{(a_{1,j})}_{n_{1,j}}=0}=\varphi_1, \quad
    \phi_{1,j}^{(a_{1,j})}(x^{(a_{1,j})}_{n_{1,j}})\Big|_{x^{(a_{1,j})}_{n_{1,j}}=x^{(a_{1,j})}_{N_{1,j}-1}}=\varphi_j, \\
    & \sum_{j=2}^4 \chi_{1,j}^{(a_{1,j})}(x^{(a_{1,j})}_{n_{1,j}})\Big|_{x^{(a_{1,j})}_{n_{1,j}}=a_{1,j}}=0, \quad
     \chi_{1,j}^{(a_{1,j})}(x^{(a_{1,j})}_{n_{1,j}})\Big|_{x^{(a_{1,j})}_{n_{1,j}}=x^{(a_{1,j})}_{N_{1,j}}}=0.
\end{split}
\end{align}

Substitution of the solution in~\eqref{sol_discrete_graph} into~\eqref{bc7} yields the secular equations
\begin{align*}
    & \sum_{j=2}^4 \frac{1}{\sin[g(a_{1,j}) x^{(a_{1,j})}_{N_{1,j}-1}]}  \left\{
           \varphi_j \cos[g(a_{1,j})a_{1,j}-r(a_{1,j}) ]
           - \varphi_1  \cos[g(a_{1,j})x^{(a_{1,j})}_{N_{1,j}-2}+r(a_{1,j}) ] \right\} = 0, \\
    & \frac{1}{\sin[g(a_{1,j}) x^{(a_{1,j})}_{N_{1,j}-1}]}\left\{
          \varphi_1  \cos[g(a_{1,j})a_{1,j}-r(a_{1,j}) ]
          - \varphi_j \cos[g(a_{1,j})x^{(a_{1,j})}_{N_{1,j}}-r(a_{1,j}) ] \right\} = 0.
\end{align*}

The above system of equations has a non-trivial solution, when
\begin{eqnarray}
    \sum_{j=2}^4 \frac{\sin[g(a_{1,j}) x^{(a_{1,j})}_{N_{1,j}-1}]}{\cos[g(a_{1,j}) x^{(a_{1,j})}_{N_{1,j}}-r(a_{1,j})]} = 0.
\end{eqnarray}
The secular equation can be numerically solved to obtain the energy eigenvalues of the discrete Dirac operator defined on the discrete quantum star graph. The first five nonzero eigenvalues for the chosen star graph are presented in Table~\ref{tab:eigenvalues_star}, where all edges have the same discretization step size, $a_{1,2}=a_{1,3}=a_{1,4}=a$, and the edge lengths are set as in the continuous case: $L_{1,2}=1.2, L_{1,3}=1.7, L_{1,4}=2.1$. In Table~\ref{tab:eigenvalues_star}, columns 2 through 6 show the nonzero eigenvalues computed for different values of the step size $a$ on the star graph.

\section{Conclusion}\label{sec_conclusion}
In this work, we studied the one-dimensional time-independent discrete Dirac equation on finite discrete chain and graphs with discrete edges. Exact analytical solutions for the eigenstates and the corresponding energy spectrum were derived, and their convergence to the continuum limit was explicitly demonstrated. We presented a table comparing discrete eigenvalues with those from the continuous case, confirming the consistency of our discrete model.

Beyond the simple interval, we extended our approach to quantum graphs with arbitrary topology in both continuous and discrete formulations by applying continuity and current conservation conditions at the vertices. In particular, we analyzed the star graph with three edges as a concrete example. The resulting eigenvalue problem was solved explicitly, and the discrete model was shown to reproduce the spectral features of the continuous graph problem in the limit of small step size.

These results illustrate that the discrete Dirac equation on graphs provides a robust and flexible tool for modeling relativistic quantum particles on branched low-dimensional structures, including confined systems and complex graph networks. This approach enables systematic treatment of boundary and matching conditions across a wide range of topologies, offering insights relevant for nanostructures, quantum wire networks, and other condensed matter systems.

Future directions may include the incorporation of external potentials, position-dependent mass terms, or interaction effects, as well as studies of time-dependent dynamics and numerical simulations of quantum transport on large or disordered graphs.

\backmatter

\noindent\textbf{Dedication:} 
Dedicated to the memory of Professor Franciszek Hugon Szafraniec, a great mathematician and for us always full of inspiration.

\bmhead{Acknowledgements}
We acknowledge the funding provided by the Grant of the Innovations Agency under the Ministry of Higher Education, Research and Innovations (FL-8824063336).

\section*{Declarations}

\begin{itemize}
\item Funding: This work is supported by European Union's Horizon 2020 research and innovation programme under the Marie Sklodowska-Curie grant agreement ID: 873071, project SOMPATY (Spectral Optimization: From Mathematics to Physics and Advanced Technology); by the grant of the Innovation Development Agency of the Republic of Uzbekistan (Ref. No. F-2021-440); and by the Grant of the Innovations Agency under the Ministry of Higher Education, Research and Innovations of the Republic of Uzbekistan (FL-8824063336).
\item Conflict of interest/Competing interests: The authors declare no competing interests.
\item Ethics approval and consent to participate: Not applicable.
\item Consent for publication: Not applicable.
\item Data availability: Not applicable.
\item Materials availability: Not applicable.
\item Code availability: Not applicable.
\item Author contribution: Not applicable.
\end{itemize}


\begin{thebibliography}{99}

\bibitem{Alonso1} V. Alonso, S. De Vincenzoza, L. Mondino, Eur. J. Phys. \textbf{18}, 315--320 (1997).
\bibitem{Alonso2} V. Alonso, S. De Vincenzo, J. Phys. A: Math. Gen. \textbf{30}, 8573--8585 (1997).

\bibitem{d_material} T.O. Wehling, A.M. Black-Schaffer, A.V. Balatsky, Advances in Physics \textbf{63}, 1--76 (2014).
\bibitem{Boyack} R. Boyack, H. Yerzhakov, J. Maciejko, Eur. Phys. J. Spec. Top. \textbf{230}, 979--992 (2021).
\bibitem{Hasan} M.Z. Hasan, G. Chang, I. Belopolski, G. Bian, S.-Y. Xu, J.-X. Yin, Nature Reviews Materials  \textbf{6}, 784--803 (2021).
\bibitem{Rakhmanov} J. Dittrich, S. Rakhmanov, D. Matrasulov, Phys. Lett. A \textbf{503}, 129408 (2024).

\bibitem{Pauling} L. Pauling, J. Chem. Phys. \textbf{4}, 673 (1936).
\bibitem{Rud} K. Ruedenberg, C.W. Scherr, J. Chem. Phys. \textbf{21}, 1565 (1953).
\bibitem{Alex} S. Alexander, Phys. Rev. B \textbf{27}, 1541 (1985).
\bibitem{Exner1} P.Exner, P.Seba, P.Stovicek, J. Phys. A: Math. Gen. \textbf{21}, 4009 (1988).
\bibitem{Kurasov} P. Kurasov, Spectral Geometry of Graphs, Springer-Verlag, Berlin (2024).

\bibitem{Uzy1} T.Kottos, U.Smilansky, Ann. Phys. \textbf{76}, 274 (1999).
\bibitem{Kuchment04} P.Kuchment, Waves in Random Media \textbf{14}, S107 (2004).
\bibitem{Uzy2} S. Gnutzmann, U. Smilansky, Adv.Phys. \textbf{55}, 527 (2006).
\bibitem{Gaspard} N. Goldman, P. Gaspard, Phys. Rev. B \textbf{77}, 024302 (2008).
\bibitem{Exner15} P. Exner, H. Kovarik, {\it Quantum waveguides} (Springer, 2015).
\bibitem{Grisha} G. Berkolaiko, P. Kuchment, {\it Introduction to Quantum Graphs, Mathematical Surveys and Monographs} AMS (2013).
\bibitem{Barra} F. Barra and P. Gaspard, J. Statist. Phys. \textbf{101}, 283 (2000).
\bibitem{Uzy3} S. Gnutzmann, J.P. Keating, F. Piotet, Ann.Phys. \textbf{325}, 2595 (2010).
\bibitem{Mugnolo} D. Mugnolo. {\it Semigroup Methods for Evolution Equations on Networks}. Springer-Verlag, Berlin, (2014).
\bibitem{Uzy4} S. Gnutzmann, H. Schanz, U. Smilansky, Phys. Rev. Lett. \textbf{110}, 094101 (2013).
\bibitem{Bolte1} J. Bolte, G. Garforth, J. Phys. A: Math. Theor. \textbf{50}, 105101 (2017).
\bibitem{Hul} O. Hul et al, Phys. Rev. E \textbf{69}, 056205 (2004).

\bibitem{Kost} V. Kostrykin, R. Schrader, J. Phys. A: Math. Gen. \textbf{32}, 595 (1999).
\bibitem{Bolte} J. Bolte, J. Harrison, J. Phys. A: Math. Gen. \textbf{36}, 2747 (2003).

\bibitem{Bolte2} J. Bolte, J. Harrison, J. Phys. A: Math. Gen. \textbf{36}, L433 (2003).
\bibitem{Harrison} J. Harrison, T. Weyand, and K. Kirsten, J. Math. Phys. \textbf{57}, 102301 (2016).

\bibitem{Dirac_tbc2} J. Yusupov, K. Sabirov, D. Matrasulov, Phys. Rev. E \textbf{104}, 014219 (2021).
\bibitem{Dirac_tbc1} J.R. Yusupov, K.K. Sabirov, Q.U. Asadov, M. Ehrhardt, D.U. Matrasulov, Phys. Rev. E \textbf{101}, 062208 (2020).

\bibitem{Behrndt} J. Behrndt, M. Holzmann, C.S.-Landauer,  G. Stenzel, Reviews in Mathematical Physics \textbf{36}, 2350036 (2024).
\bibitem{Borrelli} W. Borrelli, R. Carlone, L. Tentarelli, In: Michelangeli, A. (eds) Mathematical Challenges of Zero-Range Physics. Springer INdAM Series, \textbf{42}, 81--104 (2020).

\bibitem{Bulla} W. Bulla, T. Trenkler, J. Math. Phys. \textbf{31}, 1157 (1990).

\bibitem{DDE} N. Coskun, N. Yokus, Adv. Differ. Equ. \textbf{2020}, 409 (2020).
\bibitem{DDE2} Rafael G. Campos, L.O. Pimentel, Phys. Lett. A \textbf{266}, 98--105 (2000).
\bibitem{DDE3} K.M. Schmidt, T. Umeda, Anal. Math. Phys. \textbf{13}, 46 (2023).
\bibitem{S0}
F. Philipp, F.H.  Szafraniec, C. Trunk, 
J. Funct. Anal. 260 (2011), 1045-1059. 
\bibitem{S1} F.H. Szafraniec,
Commun. Math. Phys. 210 (2000), 323–334.
\bibitem{S2}
J. Stochel, F.H. Szafraniec, 
Glasg. Math. J. 44 (2002), 137–147.
\bibitem{S3}
F.H. Szafraniec, 
Contemp. Math. 212 (1998), 269–276.
\bibitem{S4}
F.H. Szafraniec, 
Rep. Math. Phys. 59 (2007), 401–408.
\bibitem{S5}
F.H. Szafraniec, 
Banach Center Publications 78 (2007), 293–307.
\bibitem{S6}
F.H. Szafraniec, 
Rep. Math. Phys. 53 (2004), 393–400.
\bibitem{DSE} M. Akramov, C. Trunk, J. Yusupov, D. Matrasulov, EPL \textbf{147}, 62001 (2024).
\bibitem{Weidmann1} J. Weidmann, Lineare Operatoren in Hilbertr\"aumen. Teil II: Anwendungen, Vieweg-Teubner Verlag Wiesbaden (2003).
\bibitem{Weidmann2} J. Weidmann, Spectral Theory of Ordinary Differential Operators, Springer-Verlag Berlin Heidelberg (1987).


\end{thebibliography}
\end{document}